\documentstyle[12pt]{article}

\newcounter{saveeqn}

\newtheorem{proposition}{Proposition}[section]

\newtheorem{theorem}[proposition]{Theorem}
\newtheorem{corollary}[proposition]{Corollary}
\newtheorem{definition}[proposition]{Definition}
\newtheorem{proofs}{Proof}

\newenvironment{proof}{\begin{sloppypar}\begin{proofs}\rm}{\hspace*{\fill}
$\Box$ \end{proofs} \end{sloppypar}}
\newtheorem{lempro}{Proof of the Lemma}

\newtheorem{remar}{Remark}

\newtheorem{remars}{Remarks}

\newtheorem{conse}{Consequences}

\newtheorem{exams}{Examples}

\newfont{\BigBbb}{msbm10 at 12pt}
\newcommand{\bigBbb}[1]{\mbox{\BigBbb #1}}
\newfont{\SBbb}{msbm8}

\newfont{\euft}{eufm10 at 12pt}

\newcommand{\diff}{{\rm d}}

\newcommand{\C}{\bigBbb{C}}

\newcommand{\varDelta}{{\mit\Delta}}

\begin{document}

\begin{titlepage}

\hfill LMU-TPW 98-16
\vspace*{5ex}
\begin{center}
{\Large
{\bf Riemann--Hilbert Problems for the Ernst Equation and Fibre Bundles}\\[4ex]
\large C.~Klein\\
{\em Institut f\"ur Theoretische Physik, Universit\"at T\"ubingen,\\
Auf der Morgenstelle 14, 72076 T\"ubingen, Germany}\\[2ex]
O.~Richter\\
{\em Sektion Physik der Universit\"at M\"unchen,\\
Theresienstra{\ss}e 37, 80333 M\"unchen, Germany}\\[4ex]
November 12, 1998}%
\end{center}
\vspace{3ex}
\begin{abstract}
Riemann--Hilbert techniques are used in the theory of completely integrable
differential equations to generate solutions that contain a free function 
which can be used at least in principle to solve initial or boundary 
value problems. The solution of a boundary value problem is thus reduced to 
the identification of the jump data of the Riemann--Hilbert problem from 
the boundary data. But even if this can be achieved, it is very difficult 
to get explicit solutions since the matrix 
Riemann--Hilbert problem is equivalent 
to an integral equation. In the case of the Ernst equation (the stationary 
axisymmetric Einstein equations in vacuum), it was shown in a previous work 
that the matrix problem is gauge equivalent to a scalar problem on a Riemann 
surface. If the jump data of the original problem are rational functions, 
this surface will be compact which makes it possible to give explicit 
solutions in terms of hyperelliptic theta functions. In the present work, 
we discuss Riemann--Hilbert problems on Riemann surfaces in the framework 
of fibre bundles.  This makes it possible to treat the compact and 
the non-compact case 
in the same setting and to apply general existence theorems. 
\end{abstract}
\vspace{3ex}
\centerline{PACS numbers: 04.20.Jb, 02.10.Rn, 02.30.Jr}
\end{titlepage}

\section{Introduction}\label{sec1}

Riemann--Hilbert techniques provide a powerful tool if one wants to solve 
initial or boundary value problems for completely integrable differential 
equations. They are used to generate solutions with a prescribed 
singularity structure that contain a free function. In the case of a 
boundary value problem, this function has to be chosen in a way that the 
solution takes the prescribed values at the given boundary, and similarly 
for initial value problems. Whereas this does not pose any problems in 
principle, there is little hope in practice to get explicit solutions to 
boundary value problems in this way. The reason for this is that the free 
functions (the `jump data' of the Riemann--Hilbert problem) enter  the 
solutions of the differential equation as part of an integral 
equation from which the solutions have to be constructed. This implies that 
one has to solve this integral equation first on the boundary where the 
boundary data are prescribed in order to determine the jump data. In a second 
step, one has to solve the integral equation with the then fixed jump 
data in the region under consideration. In general, both steps cannot be 
done explicitly.

In the case of the stationary axisymmetric Einstein equations, the 
situation is however different. The typical problem one has to consider 
there is the exterior of a relativistic star or a galaxy. Within these 
models, the matter leads to boundary conditions for the vacuum equations if 
a solution of the Einstein equations in the matter region 
is known. Special cases are 
two-dimensionally extended matter distributions where the field equations 
reduce to ordinary differential equations, e.g.\ disks that are 
discussed in astrophysics as models for certain galaxies. Thus one has to 
consider boundary value problems for the vacuum equations in the case of 
compact matter distributions where the 
boundary is the surface of the matter, and where the boundary data follow 
from the metric functions in the matter region. Since the vacuum equations 
are equivalent to a single complex differential equation, the Ernst 
equation \cite{ernst}, which is completely integrable 
\cite{maison,belza,neugebauer}, one can use Riemann--Hilbert techniques 
to solve the resulting boundary value problems. 

In a previous work 
\cite{prd1}, it was shown that Riemann--Hilbert problems for the Ernst 
equation with analytic 
jump functions are gauge equivalent to a scalar problem on a Riemann 
surface. In the case of rational jump data, this surface is compact which 
makes it possible to give explicit solutions to the Ernst equation in terms 
of hyperelliptic theta functions. Thus it is not necessary to consider 
integral equations in this case. The physical properties of the 
resulting class of solutions were 
discussed in \cite{prl} where it was shown that the solutions can 
have the expected regularity properties and asymptotic behaviour. In the 
present article, we discuss the Riemann--Hilbert problems on Riemann surfaces 
in the framework of fibre bundles which makes it possible to treat the 
case of compact and non compact Riemann surfaces within the same setting. 
Using a theorem of R\"ohrl \cite{roehrl}, we obtain an existence proof for 
the solutions to the Riemann--Hilbert problems. In the case of non compact 
Riemann surfaces, the constructed bundles are trivial due to Grauert's theorem 
\cite{grauert1}. 

The paper is organized as follows: In section 2, we recall that the Ernst 
equation can be treated 
as the integrability condition for an overdetermined linear 
differential system, for which we formulate the matrix Riemann--Hilbert 
problem. As an example we consider the scalar problem in the complex plane 
which can be solved with the help of the Plemelj formula \cite{musk}. 
The matrix problem for the Ernst equation is equivalent to an integral 
equation which cannot be solved explicitly in general. Explicit solutions can in 
general only be obtained if the jump matrix is diagonal thus implying that the resulting
solutions are static. In this particular case the Ernst equation reduces to the 
axisymmetric Laplace equation. By the 
help of gauge transformations of the linear system of the Ernst equation, we are able to 
transform the matrix problem to a scalar one on a Riemann surface in 
section 3. We discuss the relation between fibre bundles and Riemann--Hilbert 
problems on Riemann surfaces. These results are used to prove 
the existence for the solutions of the original problem. In the 
case of non compact Riemann surfaces, Grauert's 
theorem \cite{grauert1} implies that the constructed bundles are trivial.
On compact surfaces, we recover the explicit solutions in 
terms of theta functions.

\section{The Riemann--Hilbert Problem for the Ernst Equation}\label{sec2}
\setcounter{equation}{0}

It is well known that the vacuum metric in the  stationary and axisymmetric 
case can be written in the Weyl--Lewis--Papapetrou form (see~\cite{exac})
\begin{equation}
	\diff s^2 =-{\rm e}^{2U}(\diff t+a\diff \phi)^2+{\rm e}^{-2U}
	\left({\rm e}^{2k}(\diff \rho^2+
	\diff \zeta^2)+
	\rho^2\diff \phi^2\right),
	\label{vac1}
\end{equation}
where $\rho$ and $\zeta$ are Weyl's canonical coordinates and 
$\partial_{t}$ and $\partial_{\phi}$ are the two commuting  
(asymptotically) timelike respectively spacelike Killing vectors. In this case, 
the field equations are equivalent to the Ernst equation for the potential $f$ 
where $f={\rm e}^{2U}+{\rm i}b$, and where the real function $b$ 
is related to the metric functions via
$b_{z}=-({\rm i}/\rho){\rm e}^{4U}a_{,z}$.
Here the complex variable $z$ stands for $z=\rho+{\rm i}\zeta$. With these 
settings, the Ernst equation reads
\begin{equation}\label{3.3}
f_{z\bar{z}}+\frac{1}{2(z+\bar{z})}(f_{\bar{z}}+f_z)=\frac{2 }{f+\bar{f}}
f_z f_{\bar{z}}\label{vac10}\enspace,
\end{equation}
where a bar denotes complex conjugation in $\bar{\C}$.
With a solution of the Ernst equation, 
the metric function $U$ follows directly from the definition of the Ernst 
potential whereas $a$ and $k$ can be obtained from $f$ via quadratures.

The importance of the formulation of the field equations in terms of the 
Ernst potential arises from the fact that the Ernst equation is completely 
integrable, see~\cite{maison,belza,neugebauer}. This means that the latter can 
be treated as the integrability condition of an overdetermined linear 
differential system that contains an additional complex parameter, the so 
called spectral parameter, that reflects an underlying symmetry of the Ernst 
equation. We use the linear system for the $2\times2$--matrix $\Phi$ 
of \cite{neugebauer},
\begin{eqnarray}
	\Phi_{z}(K,z,\bar{z}) & = & \left\{\left(
	\begin{array}{cc}
		N & 0  \\
		0 & M
	\end{array}
	\right)+\frac{K-{\rm i}\bar{z}}{\mu_0}\left(
	\begin{array}{cc}
		0 & N  \\
		M & 0
	\end{array}
	\right)\right\}\Phi(K,z,\bar{z}),
	\label{lin1} \\
	\Phi_{\bar{z}}(K,z,\bar{z}) & = & \left\{\left(
	\begin{array}{cc}
		\bar{M} & 0  \\
		0 & \bar{N}
	\end{array}
	\right)+\frac{K+{\rm i}z}{\mu_0}\left(
	\begin{array}{cc}
		0 & \bar{M}  \\
		\bar{N} & 0
	\end{array}
	\right)\right\}\Phi(K,z,\bar{z}).
	\label{lin2}
\end{eqnarray}
Here the spectral parameter $K$ resides on a family of Riemann surfaces 
${\cal L}(z,\bar{z})$ parametrized by the physical coordinates $z$ 
and $\bar{z}$ and given by $\mu_0^2(K)=(K-{\rm i}\bar{z})(K+{\rm i}z)$. 
A point on ${\cal L}$ is 
denoted by $P=(K,\mu_0(K))$ with $K\in\bar{\C}$. 
The functions $M$ and $N$ depend only on $z$ and $\bar{z}$ but not on $K$, 
and have the form
\begin{equation}
M=\frac{f_{z}}{f+\bar{f}}\enspace, \quad 
N=\frac{\bar{f}_{z}}{f+\bar{f}}\enspace,
\label{lin3.1}
\end{equation} 
where $f$ is again the Ernst potential.  

To construct  solutions to the Ernst equation with the help of 
the above linear system, one 
investigates the singularity structure of the matrices $\Phi_{z}\Phi^{-1}$ and 
$\Phi_{\bar{z}}\Phi^{-1}$ with respect to the spectral parameter and infers 
a set of conditions for the matrix $\Phi$ that satisfies the linear system 
(\ref{lin1}) and (\ref{lin2}). 
This is done (see e.g.\ \cite{korot1}) in 
\begin{theorem}\label{thm1}
Let $\Phi$ be subject to the following conditions:\\
I. $\Phi(P)$  is holomorphic and invertible at the branch 
points $P_0=-{\rm i}z$ and $\bar{P}_0$ such that the logarithmic derivative
$\Phi_{z}\Phi^{-1}$ diverges as $(K+{\rm i}z)^{\frac{1}{2}}$ at $P_0$ and 
$\Phi_{\bar{z}}\Phi^{-1}$ as $(K-{\rm i}\bar{z})^{\frac{1}{2}}$ at $\bar{P}_0$.\\
II. All singularities of $\Phi$ on ${\cal L}$ (poles, essential 
singularities, zeros of the determinant of $\Phi$, branch cuts and branch 
points) are regular which means that the logarithmic derivatives 
$\Phi_{z}\Phi^{-1}$ and $\Phi_{\bar{z}}\Phi^{-1}$ are holomorphic there.\\ 
III. $\Phi$ is subject to the reduction condition
\begin{equation}
	\Phi(P^{\sigma}) = \sigma_3 \Phi(P) \sigma_1
	\label{lin7}
\end{equation}
where $\sigma$ is the involution on ${\cal L}$ that interchanges the sheets, 
and $\sigma_1$ and $\sigma_3$ are Pauli matrices.\\
IV. The normalization and reality condition
\begin{equation}
	\Phi(P=\infty^+)=\left(
	\begin{array}{rr}
		\bar{f} & 1  \\
		f & -1
	\end{array}
	\right)
	\label{lin9}
\end{equation}
is fulfilled. Then the function $f$ in (\ref{lin9}) is a solution to the Ernst equation.
\end{theorem}
This theorem has the following 
\begin{corollary}
Let $\Phi(P)$ be a matrix subject to the conditions of  Theorem 2.1
and $C(K)$ be a $2\times2$--matrix that only depends on  $K\in \bar{\C}$ 
with the properties (the $\alpha_i$ are scalar functions)
\begin{eqnarray}
C(K) & = & \alpha_1(K) \hat{1}+\alpha_2(K) \sigma_1,\nonumber \\
	\alpha_1(\infty) &=  &1,\quad \alpha_2(\infty)=0.
	\label{c}
\end{eqnarray}
Then the matrix $\Phi'(P)=\Phi(P) C(K)$ also satisfies the conditions of 
Theorem 2.1 and $\Phi'(\infty^+)=\Phi(\infty^+)$.
\end{corollary}
In other words: matrices $\Phi$ which are related through the 
multiplication  from the right by a matrix $C$ of the above form lead to 
the same Ernst potential though their singularity structure may be vastly 
different (the functions $\alpha_i$ need not be holomorphic). Therefore 
this multiplication is called a gauge transformation.

Theorem \ref{thm1} can be used to construct solutions to the 
Ernst equation by determining the structure and the singularities of 
$\Phi$ in accordance with the conditions I to IV. In the present paper, we 
will concentrate on the Riemann--Hilbert problem for the Ernst equation 
which can be formulated in the following form:
Let $\Gamma$ be a set of (orientable piecewise smooth) 
contours $\Gamma_k \subset {\cal L}$ ($k=1,\dots,l$)  such that 
with $P\in \Gamma$ also $\bar{P}\in
\Gamma$ and $P^{\sigma}\in \Gamma$. Let  ${\cal G}_k(P)$ be
matrices on $\Gamma_k$ with analytic components and 
nonvanishing determinant subject to the reality condition ${\cal 
G}_{ii}(\bar{P})=\bar{{\cal G}}_{ii}(P)$ for the diagonal elements, and 
${\cal G}_{ij}(\bar{P})=-\bar{{\cal G}}_{ij}(P)$ for the offdiagonal 
elements. We define $\gamma(t,\Gamma_j)=1$ if $t\in \Gamma_j$ and $0$ 
otherwise and ${\cal G}=\sum_{k=1}^l \gamma(t,\Gamma_k){\cal G}_k$. 
Let ${\cal G}(P^{\sigma})=\sigma_1{\cal G}(P)\sigma_1$. 
Both $\Gamma$ and ${\cal G}$ have to be 
independent of $z$, $\bar{z}$. The matrix $\Phi$ has to be everywhere 
regular except at the contour $\Gamma$ where the boundary values 
on both sides of the contours (denoted by $\Phi_{\pm}$) are related via
\begin{equation}
	\left.\Phi_- (P)=\Phi_+ (P) {\cal G}_i(P)\right|_{P\in \Gamma_i}
	\label{lin6}.
\end{equation}

It may be easily checked that a matrix $\Phi$ constructed in this way
satisfies the conditions of Theorem~\ref{thm1}. Furthermore, it can be seen from 
the Theorem that 
the only possible singularities of the Ernst potential can occur where the 
conditions are not satisfied, i.e.\ where $\Phi$ cannot be normalized or 
where $P_0$ coincides with one of the singularities of $\Phi$, in our case 
the contour $\Gamma$. The latter makes the Riemann--Hilbert problem very 
useful if one wants to solve boundary value problems for the Ernst 
equation: choose the contour $\Gamma$ in a way that $P_0\in \Gamma$ just 
corresponds to the contour in the meridian ($z,\bar{z}$)-plane where the 
boundary values are prescribed. The Ernst potential will in general 
not be continuous 
at this contour, but its boundary values will be bounded. Notice however 
that the Ernst potential will not always be singular if $P_0$ 
coincides with a singularity of $\Phi$ since the latter may be e.g.\ pure gauge. 
Theorem~\ref{thm1} merely ensures that the solution will be regular at all 
other points. 

To solve Riemann--Hilbert problems, one basically uses the same methods as 
in the simplest case, the problem in the complex plane for a scalar 
function $\psi$, see e.g.~\cite{zverovich1}. If $\Gamma_K$ is a simply connected 
closed smooth contour and 
$G$ a nonzero H\"older continuous function on $\Gamma_K$ in $\C$, the 
function  $\psi$ that is holomorphic except at the contour $\Gamma_K$ where 
\begin{equation}
	\psi_-=\psi_+ G
	\label{rh.1}
\end{equation}
is obviously given by the Cauchy integral,
\begin{equation}
	\psi(K)=F(K)\exp\left(\frac{1}{2\pi {\rm i}}
	\int_{\Gamma_K}^{}\frac{\ln G dX}{X-K}\right),
	\label{rh.2}
\end{equation}
where $F(K)$ is an arbitrary holomorphic function, and where the principal 
value of the logarithm has to be taken. The well--known analytic properties 
of the Cauchy integral ensure that condition (\ref{rh.1}) is satisfied. 
Formula (\ref{rh.2}) shows that the solution to a Riemann--Hilbert problem 
of the above form is only determined up to a holomorphic function. Since $F$ 
is holomorphic, the solution will be uniquely determined (due to Liouville's
theorem) by a normalization condition $\psi(\infty)=\psi_0$ for 
$\infty\notin\Gamma_K$. 
Uniqueness is lost if one allows for additional poles  since $F$ in 
(\ref{rh.2}) has then to be replaced by a meromorphic function. We note 
that the above conditions on the contour may be relaxed:  $\Gamma$ may 
consist of a set of piecewise smooth orientable contours which are not 
closed. The Plemelj formula, see e.g.\ \cite{musk}, assures that formula 
(\ref{rh.2}) still gives the solution to (\ref{rh.1}). A normalization 
condition will however only establish uniqueness of the solution if $G=1$ 
at the endpoints of $\Gamma$, i.e.\ if the index of the problem 
(\ref{rh.1}) vanishes.

Riemann--Hilbert problems on the sphere ${\cal L}$ which occur in the case of the 
Ernst equation can be treated in much 
the same way as the problems in the complex plane. The basic building 
block for the solutions is the differential of the third kind 
$d\omega_{K^+K^-}(X)$ that 
corresponds to the differential $dX/(X-K)$ in the complex plane, 
i.e.\ a differential that can 
be locally written as $F(X,K)dX$ where $F(X,K)$ is holomorphic except 
for $X=K^{\pm}$ where the residue is $\pm1$. If we make the ansatz 
\begin{equation}
	\Phi-\Phi_0=\frac{1}{2\pi {\rm i}}\int_{\Gamma}^{}\chi(X)d\omega_{K^+K^-}(X)
	\label{rh3},
\end{equation}
where the $2\times 2$-matrix $\chi$ is given by 
$\chi=\sum_{k=1}^{l}\gamma(t,\Gamma_k)\chi_k$ and where $\Phi_0$ is holomorphic,
we get for (\ref{lin6}) at the contour $\Gamma$
with the Plemelj formula $\Phi^{\pm}=\pm\frac{1}{2} \chi+\frac{1}{2\pi {\rm 
i}}\int_{\Gamma}^{}\chi(X)d\omega_{K^+K^-}(X)$. Thus the Riemann--Hilbert 
problem (\ref{lin6}) is equivalent to the integral equations (at $\Gamma$) 
\begin{equation}
	\frac{1}{2}\chi({\cal G}+1)+\frac{1}{2\pi {\rm 
	i}}\int_{\Gamma}^{}\chi(X)d\omega_{K^+K^-}(X) ({\cal G}-1)=0\enspace.
	\label{rh.3}
\end{equation}

For simplicity, we will only consider the case where the projection of the 
contour $\Gamma$ into the complex plane
has a simply connected component $\Gamma_K$. 
In \cite{prd1} we have shown that the 
general problem (\ref{lin6}) is gauge equivalent to 
a problem with ${\cal G}_1=\left(
\begin{array}{cc}
	\alpha & 0  \\
	\beta & 1
\end{array}
\right)$ on $\Gamma_1$ which is the contour in the $+$ sheet. This gauge 
transformation does not change the singularity structure of $\Phi$ (i.e.\ 
$\Phi$ will only be singular at $\Gamma$). The 
reduction condition III of Theorem 2.1 implies that ${\cal G}_2=\left(
\begin{array}{cc}
	1 & \beta  \\
	0 & \alpha
\end{array}\right)$ on the contour $\Gamma_2$ in the $-$-sheet. Because of the 
reality conditions for ${\cal G}$, this implies that solutions to 
the Ernst equation that follow from (\ref{lin6}) contain two real valued
functions which correspond to $\alpha$ and $\beta$. The reduction and 
reality properties of $\Phi$ (see Theorem 2.1) make it possible to 
consider only one component of the matrix, e.g.\ $\Phi_{12}$ from which 
the Ernst potential follows as $f=\Phi_{12}(\infty^+)$. With 
$\Phi_{12}=\psi_0+\frac{1}{4\pi {\rm 
i}}\int_{\Gamma_1}^{}\frac{\mu_0(K)+\mu_0(X)}{2\mu_0(X)(X-K)}Z(X,z,\bar{z})dX$ 
we obtain the Ernst potential for given $\alpha$ and $\beta$ where $Z$ is 
the solution of the integral equation 
\begin{equation}
	-\frac{\alpha+1}{2}Z=\frac{\alpha-1}{4\pi{\rm i}}\int_{\Gamma_1}^{}
	\frac{\mu_0(K)+\mu_0(X)}{\mu_0(X)(X-K)}Z(X,z,\bar{z})dX-
	\frac{\beta}{4\pi{\rm i}}\int_{\Gamma_1}^{}
	\frac{\mu_0(X)-\mu_0(K)}{\mu_0(X)(X-K)}Z(X,z,\bar{z})dX,
	\label{rh.4}
\end{equation}
and where $\psi_0$ follows from the normalization condition 
$\Phi_{12}(\infty^-)=1$.

Explicit solutions can in general only be obtained for diagonal ${\cal 
G}$, i.e.\ for $\beta=0$. In this case we get with the above formulas
$f=\bar{f}={\rm e}^{2U}$ with 
\begin{equation}
	U=-\frac{1}{4\pi {\rm i}}\int_{\Gamma_1}^{}\frac{\ln 
	\alpha}{\sqrt{(K-\zeta)^2+\rho^2}}dK
	\label{rh.5}.
\end{equation}
Thus all solutions are real in this case which implies that 
they belong to the static Weyl class. Since the Ernst equation reduces to 
the axisymmetric Laplace equation for $U$ if $f$ is real, the function $U$ 
in (\ref{rh.5}) solves the Laplace equation. In fact one can show that the 
contour integral there is equivalent to the Poisson integral with a 
distributional density. It can also be directly seen from the expression 
(\ref{rh.5}) that the dependence on the physical coordinates $\rho$ and 
$\zeta$ enters through the branch points of the family of surfaces ${\cal L}$.

\section{Riemann-Hilbert Problems on Riemann Surfaces and Vector 
Bundles}\label{sec3}
\setcounter{equation}{0}

In the previous section we recalled that matrix Riemann--Hilbert 
problems cannot be solved in general. 
Only for particular cases one can find an explicit form of the solution. 
In \cite{prd1} we have shown that in the context of the Ernst
equation it is, however, possible to go one step further if one drops the 
condition that the gauge transformed matrix $\Phi'$ has the same singularity 
structure as the original matrix $\Phi$ in (\ref{lin6}). It was furthermore
shown there that the Riemann--Hilbert problem (\ref{lin6}) is gauge equivalent
to a problem with diagonal matrix ${\cal G}'={\rm diag}(G,1)$ on a two sheeted
covering $\hat{\cal L}$ of ${\cal L}$, given by an equation of the form
\begin{equation}\label{3.1}
\hat{\mu}^2(K)=F(K)\,H(K)\enspace,
\end{equation}
where $F(K)$ and $H(K)$ are holomorphic functions. They follow from 
the jump matrix ${\cal G}$ via
\begin{equation}\label{gauge10}
\frac{F(K)}{H(K)}=\frac{({\cal G}_{11}-{\cal G}_{12}
+{\cal G}_{21}-{\cal G}_{22})({\cal G}_{11}-{\cal G}_{12}
-{\cal G}_{21}+{\cal G}_{22})}{({\cal G}_{11}+{\cal G}_{12}-{\cal G}_{21}-{\cal G}_{22})
({\cal G}_{11}+{\cal G}_{12}
+{\cal G}_{21}+{\cal G}_{22})}\enspace,
\end{equation}
whereas the analytic jump function $G$ can be expressed via the components of the 
original jump matrix ${\cal G}$ by
\begin{equation}\label{3.2}
\frac{G+1}{G-1}=
\sqrt{\frac{({\cal G}_{11}-{\cal G}_{12}-{\cal G}_{21}+{\cal G}_{22})
({\cal G}_{11}+{\cal G}_{12}+{\cal G}_{21}+{\cal G}_{22})}
{({\cal G}_{11}-{\cal G}_{12}+{\cal G}_{21}-{\cal G}_{22})
({\cal G}_{11}+{\cal G}_{12}-{\cal G}_{21}-{\cal G}_{22})}}\enspace.
\end{equation}
By definition a Riemann surface is given by an equation of the form 
$f(K,\mu)=0$, where $f(K,\mu)$ is an analytic function of $K$ and $\mu$. If 
$f(K,\mu)$ is a polynomial in both variables one speaks of a compact Riemann 
surface. Therefore, for our purposes it is sufficient to solve a scalar
Riemann--Hilbert problem on a four sheeted Riemann surface which is, dependent
on the initial variables ${\cal G}_{ij}$, either compact or non
compact (if the components of ${\cal G}$ are rational functions, the 
surface will be compact). 

In the mentioned paper we restricted ourselves to the case of
$\hat{\cal L}$ being compact with genus $g$, where it is possible to 
give explicit solutions to the Riemann--Hilbert problem in terms of theta 
functions. Here we give a characterization of solutions to
Riemann--Hilbert problems in terms of fibre bundle theory, which allows for a
treatment of both the compact and non compact case.

It is a well known fact, see~\cite{rodin}, that there is a relation between 
Riemann-Hilbert problems on Riemann surfaces and holomorphic vector bundles 
over them. To make the paper self consistent, we begin with a brief 
introduction into fibre bundle theory, see~\cite{kobayashi1,nakahara1}.

Let us recall what a {\em differentiable fibre bundle} is.
\begin{definition}
A {\em differentiable fibre bundle} $(E,M,F,G,\pi)$ is a 5-tuple consisting
of\\
I. A differentiable manifold $E$ -- the so called {\em total space}.\\
II. A differentiable manifold $M$ -- the so called {\em base space}.\\
III. A differentiable manifold $F$ -- the {\em fibre} or {\em typical
fibre}.\\
IV. A surjective map $\pi:E\to M$, called {\em projection}. The
inverse image $\pi^{-1}(p)\equiv F_p\simeq F$ of $p\in M$ is called the {\em
fibre} at $p$.\\
V. A Lie group $G$ -- the {\em structure group} -- acting on $F$ on
the left.\\
VI. An open covering $\left\{U_i\right\}$ of $M$ together with
diffeomorphisms $\phi_i:U_i\times F\to\pi^{-1}(U_i)$, such that
$\pi\circ\phi_i(p,f)=p$ ($p\in U_i$, $f\in F$). We call $\phi_i$ the 
{\em local trivialisation} since
$\phi_i^{-1}$ maps $\pi^{-1}(U_i)$ onto the direct product $U_i\times F$.\\
VII. If we set $\phi_{i,p}(f)\doteq\phi_i(p,f)$ then $\phi_{i,p}:F\to
F_p$ is a diffeomorphism. If $U_i\cap U_j\not=\emptyset$ we require that
$t_{ij}(p)\doteq\phi_{i,p}^{-1}\phi_{j,p}:F\to F$ be an element of the
structure group $G$. Then $\phi_i$ and $\phi_j$ are related by a smooth map
$t_{ij}:U_i\cap U_j\to G$ as
\begin{equation}\label{1.1}
\phi_j(p,f)=\phi_i(p,t_{ij}(p)f)\enspace.
\end{equation}
We call the $\left\{t_{ij}\right\}$ the {\em transition functions}.
\end{definition}
Let us say a few words about this definition. If we take a chart $U_i$ of the
base space $M$ then $\pi^{-1}(U_i)$ is diffeomorphic to $U_i\times F$ with
diffeomorphism $\phi_i^{-1}:\pi^{-1}(U_i)\to U_i\times F$. If the intersection
$U_i\cap U_j\not=\emptyset$, there are two maps $\phi_i$ and $\phi_j$ on this
intersection. Let $u\in E$ such that $\pi(u)=p\in U_i\cap U_j$. We then have
\begin{eqnarray*}
\phi_i^{-1}(u)&=&(p,f_i)\\
\phi_j^{-1}(u)&=&(p,f_j)\enspace.
\end{eqnarray*}
There is a map $t_{ij}:U_i\cap U_j\to G$ which relates $f_i$ and $f_j$:
$f_i=t_{ij}(p)f_j$.

The transition functions can not be chosen arbitrarily in order to be the
transition functions of a fibre bundle. They have to satisfy some consistency
conditions:
\begin{eqnarray}\label{cons1}
t_{ii}(p)&=&\mbox{identity map}\enspace,\enspace\enspace (p\in U_i)\nonumber\\
t_{ij}(p)&=&t_{ji}(p)^{-1}\enspace,\enspace\enspace 
(p\in U_i\cap U_j)\nonumber\\
t_{ij}(p)t_{jk}(p)&=&t_{ik}(p)\enspace,\enspace\enspace
(p\in U_i\cap U_j\cap U_k)\enspace.
\end{eqnarray}
A fibre bundle is called {\em trivial}, if all the transition functions can be
taken to be identity maps. A trivial bundle is of the form $E=M\times F$.

The transition functions are so important because they contain all the
information needed to construct a fibre bundle. Let us now show how starting from 
a 5-tuple ($M,\{U_i\},\{t_{ij}(p)\},F,G$) a fibre bundle over $M$ with typical fibre $F$
can be constructed. Finding the bundle means finding unique $E$, $\pi$ and $\phi_i$ 
from the above data. We define
\begin{equation}%\label{
X\equiv\bigcup\limits_i (U_i\times F)\enspace,
\end{equation}
and introduce on $X$ an equivalence relation $\sim$ as follows. We say that
$(p,f)
\in U_i\times F$ and $(q,f')\in U_j\times F$ are equivalent, $(p,f)\sim
(q,f')$, if and only if $p=q$ and $f'=t_{ij}(p)f$. The total space $E$ of the
fibre bundle is then defined 
\begin{equation}%\label{
E=X/\sim\enspace.
\end{equation}
We denote an element of $E$ by the equivalence class $[(p,f)]$. The projection
$\pi:E\to M$ is given by
\begin{equation}%\label{
\pi:[(p,f)]\to p\enspace,
\end{equation}
and the local trivialisation $\phi_i:U_i\times F\to\pi^{-1}(U_i)$ is given by
\begin{equation}%\label{
\phi_i:(p,f)\to [(p,f)]\enspace,
\end{equation}
with $p\in U_i$ and $f\in F$. The above data reconstruct the bundle $E$ uniquely. 

Let us now make the relation between Riemann--Hilbert problems on Riemann
surfaces and vector bundles on them more explicit.

\subsection{The compact case}

First we will consider the case that the Riemann surface $\hat{\cal L}$ 
obtained as the double covering via the procedure described in~\cite{prd1} of 
the Riemann surface ${\cal L}$ is compact, i.e.~we consider the 
Riemann--Hilbert problem 
\begin{equation}\label{rh1}
\phi_-(P)=\phi_+(P)\, G(P)
\end{equation}
on a compact Riemann surface. Due to the fact that on such surfaces theta 
functions are the basic building blocks for the construction of meromorphic
functions, see~\cite{mumford1}, we may express the solution to (\ref{rh1}) in
terms of these functions. But, in order to make contact with the case of 
$\hat{\cal L}$ being non compact, we describe here how a solution to 
(\ref{rh1}) is connected with some line bundle over $\hat{\cal L}$. To this end 
we want to make use of the above result 
that a vector bundle over a manifold $\hat{\cal L}$ is completely 
determined by a triple $(\hat{\cal L},\{U_i\},\{t_{ij}\})$. Let 
$\{U_i\}$ be a covering of $\hat{\cal L}$ and $\{\phi_i(P)\}$ solutions to 
(\ref{rh1}) in the domains $U_i$, different from zero. In the domains $U_i$ the
original scalar Riemann--Hilbert problem is reduced to a problem on the complex plane 
$\C$, which can be solved (see the discussion in section \ref{sec2}).

In other words, the functions $\phi_i(P)$ are non vanishing on $U_i$ and fulfil 
(\ref{rh1}) on the intersection $\Gamma\cap U_i$. If $\Gamma\cap
U_i=\emptyset$ then $\phi_i(P)$ is a holomorphic function in $U_i$. Let us now 
define some functions $t^{\pm}_{ij}(P)$ in $U_i\cap U_j$ by
\begin{equation}\label{rh2}
t^{\pm}_{ij}(P)=\frac{\phi_{j\pm}(P)}{\phi_{i\pm}(P)}\enspace,
\end{equation}
for $P\in U_i\cap U_j$. We have
\begin{equation}\label{rh21}
t^{-}_{ij}(P)=\frac{\phi_{j-}(P)}{\phi_{i-}(P)}=
\frac{\phi_{j+}(P)G(P)}{\phi_{i+}(P)G(P)}=
\frac{\phi_{j+}(P)}{\phi_{i+}(P)}=t^{+}_{ij}(P)\enspace,
\end{equation} 
i.e.~the functions $t_{ij}(P)\doteq t^{+}_{ij}(P)=t^{-}_{ij}(P)$ do not jump at 
the contour $\Gamma$. It follows immediately from this definition that the 
$t_{ij}$ fulfil the consistency conditions for transition functions 
(\ref{cons1}). Because the $\phi_i(P)$ are non vanishing the functions $t_{ij}(P)$ 
take values in $\C^{*}$ (the complex numbers different from zero). Therefore, we may 
define a complex bundle with structure group $\C^{*}$ and 
standard fibre $\C$, i.e.~a line bundle, $B_G$, over the compact Riemann
surface $\hat{\cal L}$ by the 5-tupel $(\hat{\cal L}, \{U_i\}, \{t_{ij}(P)\},\C,\C^*)$. 
In other words: for $\hat{\cal L}$ being compact we may associate to a 
Riemann--Hilbert problem (\ref{rh1}) a line bundle over this surface.

\subsection{The non compact case}

Let us now turn to the case $\hat{\cal L}$ being non compact. Contrary to the 
compact case we do not have the calculus of theta functions associated to a
Riemann surface at our disposal in order to construct meromorphic functions. 
Nevertheless, we may perform a similar construction as in the compact case and
construct a vector bundle for a Riemann--Hilbert problem given on this surface.
The remarkable point is now that this vector bundle is, due to a 
theorem by Grauert \cite{grauert1}, a trivial one. 

To be more precise, let $\hat{\cal L}$ be equipped with a covering 
${\cal N}=\{U_i,i\in I\}$, where $I$ denotes
some set of indices. Let us suppose that there exists a number $N$, the
covering constant, such that any point $P\in\hat{\cal L}$ belongs to no more
than $N$ domains of the covering, see~\cite{rodin}. We assume that the contour $\Gamma$
is compact and closed, dividing $\hat{\cal L}$ into, in general, non compact domains. 
To simplify the discussion we are looking for solutions $\phi(P)$ with finite Dirichlet 
integral, i.e.
\begin{equation}\label{rh42}
\int\limits_{\hat{\cal L}\setminus\varDelta(\Gamma)}\,
\overline{\diff\phi}\wedge\diff\phi <\,\infty
\enspace,
\end{equation}
where $\varDelta(\Gamma)$ denotes some neighbourhood of $\Gamma$. Let $\phi_i(P)$ be, as
above, a non vanishing function on $U_i$ and solving there the Riemann--Hilbert
problem (\ref{rh1}) on the intersection $\Gamma\cap U_i$. If $\Gamma\cap
U_i=\emptyset$ then $\phi_i(P)$ is a holomorphic function in 
$U_i$. Similarly to the compact case we now define functions $t^{\pm}_{ij}(P)$ 
in $U_i\cap U_j$ by
\begin{equation}\label{rh4}
t^{\pm}_{ij}(P)=\frac{\phi_{j\pm}(P)}{\phi_{i\pm}(P)}\enspace.
\end{equation}
Again, we have
\begin{equation}\label{rh41}
t^{-}_{ij}(P)=\frac{\phi_{j-}(P)}{\phi_{i-}(P)}=
\frac{\phi_{j+}(P)G(P)}{\phi_{i+}(P)G(P)}=
\frac{\phi_{j+}(P)}{\phi_{i+}(P)}=t^{+}_{ij}(P)\enspace,
\end{equation} 
for $P\in U_i\cap U_j$, i.e.~the functions $t_{ij}(P)\doteq t^{+}_{ij}(P)=t^{-}_{ij}(P)$
do not jump at $\Gamma$, analogously to the compact case. These functions also obey the
consistency conditions (\ref{cons1}) and, therefore, we may associate to the
Riemann--Hilbert problem (\ref{rh1}) a vector bundle in the same manner as in 
the compact case. But, whereas in the compact case one does not know in advance 
what the global structure of the bundle space $B_G$ looks like, we have in the
present case the following
\begin{theorem}[Grauert]
Any complex line bundle over a non compact Riemann surface is trivial.
\end{theorem}
\begin{proof}
The proof can be found in~\cite{grauert1}.
\end{proof}
From this theorem it follows that for non compact $\hat{\cal L}$ the line bundle 
$B_G$ associated the Riemann--Hilbert problem (\ref{rh1}) has the form
\begin{equation}\label{grauert1}
B_G\simeq\hat{\cal L}\times \C\enspace.
\end{equation}
To conclude, we have shown that for $\hat{\cal L}$ being non compact
(where one does not have an explicit solution of the Riemann--Hilbert problem
in terms of theta functions) there is a simple geometric characterization of it 
in terms of fibre bundles over $\hat{\cal L}$. 
Due to its local properties the bundle approach allows to reduce 
the scalar Riemann--Hilbert problems on $\hat{\cal L}$ to problems on the
complex plane $\C$ which can be explicitly solved. For non compact
Riemann surfaces $\hat{\cal L}$ the resulting total space is given as a direct
product whereas in the compact case the total space is, in general, twisted in
a non trivial manner. On the other hand, in this last case, an explicit 
formulation of the solution in terms of theta functions is possible.\\[1ex]
{\bf Acknowledgments:}\\
We thank H.~R{\"ohrl} for a helpful discussion and hints, and 
J.~Frauendiener for the careful reading of the manuscript. The authors
acknowledge support by the DFG.

\end{document}